\documentclass[twocolumn,showpacs,preprintnumbers,amsmath,amssymb]{revtex4}
\usepackage{graphicx}

\begin{document}

  \title{An Introduction to the Colloidal Glass Transition}

\author{Eric R. Weeks}
\affiliation{Department of Physics, Emory University,
Atlanta, GA 30322, USA}
\email{erweeks@emory.edu}

\date{\today}

\begin{abstract}
Colloids are suspensions of small solid particles in a liquid, and
exhibit glassy behavior when the particle concentration is high.
In these samples, the particles are roughly analogous to individual
molecules in a traditional glass.  This model system has been
used to study the glass transition since the 1980's.  In this
Viewpoint we summarize some of the intriguing behaviors of the
glass transition in colloids, and discuss open questions.
\end{abstract}

\maketitle



Glasses are an intriguing state of matter in that they share some
similarities to both liquids and solids.  Molten glass is a liquid
and can flow easily, but as it cools its viscosity rises smoothly.
In fact upon cooling by several tens of degrees, the viscosity grows
by ten to twelve orders of magnitude.  One rough definition of
when a sample becomes a glass is when its viscosity is $10^{15}$
times that of water, simply because viscosities that are any larger
become problematic to measure.  At this point the sample remains as
disordered as a liquid on the molecular scale, but macroscopically
appears solid.  This is perhaps a dissatisfying situation,
in that regular phase transitions are more obvious
and well-defined as to the precise temperatures and pressures
at which they occur.  In contrast, the temperature required to
form a glass depends on the cooling rate.  Furthermore, one can
note that if one waits decades flow can sometimes be observed
\cite{edgeworth84}, although this is not relevant for
window glass \cite{zanotto98,pasachoff98}.

Polymers easily form glasses.  In part this is because the
polymers entangle, which already causes interesting nontrivial
flow properties even when a polymer sample isn't glassy.  More to
the point, polymers generally have large viscosities and so
even slowly cooled polymers have difficulty rearranging into
a crystalline state.  Polymers with stereo-irregular chemistry
(random placement of side groups) further frustrate crystallization.
For these reasons, when cooled polymers are quite likely to be
trapped in a glassy state.  Plastic materials are polymer glasses.

In the 1980's, colloidal suspensions were introduced as model
systems which had a glass transition
\cite{lindsay82,pusey86,pusey87}.  Colloidal suspensions are
composed of small (10 nm - 10 $\mu$m radius) solid particles in a
liquid.  Their glass transition is not as a function of temperature,
but rather of concentration.  At low concentration, particles
undergo Brownian motion and diffuse through the sample freely.  At
higher concentrations, the particles pack together randomly (with a
liquid-like structure), and macroscopically the sample viscosity
grows dramatically as a function of concentration.  Below the glass
transition concentration, Brownian motion enables the sample to
equilibrate, and the sample is still considered a liquid.  Above the
glass transition concentration, equilibration is no longer possible
on experimental time scales, and macroscopically the sample has a
yield stress like a regular elastic material.

\begin{figure}
\includegraphics[width=6cm]{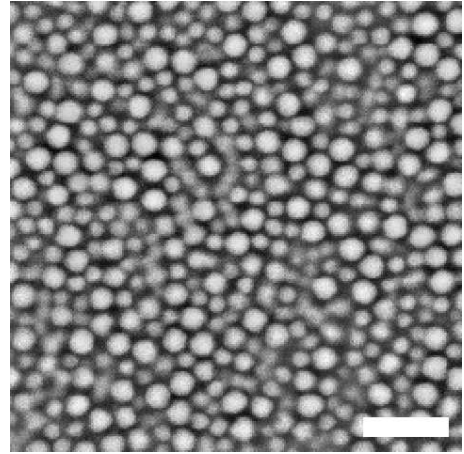}
\caption{
Confocal microscopy image of a bidisperse colloidal sample 
with particle radii 1.18~$\mu$m and 1.55~$\mu$m.  The scale bar
represents 10~$\mu$m.
Reproduced from Ref.~\cite{narumi11} with permission from the
Royal Society of Chemistry.
}
\label{narumipic}
\end{figure}

Colloidal glasses share many similarities to ``regular''
glasses.  For example, they have a strong growth of their
viscosity as the glass transition is approached \cite{cheng02};
their structure is essentially unchanged at the glass transition
\cite{vanblaaderen95}; materials become dynamically heterogeneous
as the transition is approached \cite{weeks00,kegel00}; confining
colloidal samples modifies their glass transition
\cite{nugent07prl,zhang16}.
This Viewpoint cannot describe all of the interesting glassy
phenomena that have been studied with colloidal glasses,
although the reader is invited to consult longer review articles
\cite{hunter12rpp,joshi14,pusey08,sciortino05,gokhale16aip}.   Rather,
a few representative examples will be presented below that
will highlight the advantages of colloids as a model system.
A particular advantage is that their large size makes colloids
directly observable with optical microscopy (see
Fig.~\ref{narumipic} as well as indirectly observable with light
scattering \cite{scheffold07}.

Colloidal particles interact with one
another with a variety of forces.  This can include repulsive forces
(such as electrostatic forces if the particles are charged) and
attractive forces (the van der Waals force due to fluctuating
electric dipole moments of the particles, which is quite strong at
short range).  Discussing these forces is beyond the scope of this
review, so accordingly I will focus the discussion on purely
repulsive colloidal particles.  One important category is
hard-sphere-like particles \cite{pusey86}.  Typically these are made
by suspending the particles in a solvent that matches their index of
refraction (thus reducing the van der Waals force), adding some sort
of salt (thus screening the electrostatic forces), and coating the
particles with a polymer brush layer.  This polymer brush prevents
the particles from approaching too closely, further preventing
particles from sticking together due to the attractive van der Waals
forces.  Frequently this polymer stabilizing layer is short (a
length of $10-20$ nm coating a particle of diameter $\sim 1$~$\mu$m)
\cite{pusey86,poon12} and so the
particles can be treated as hard-sphere-like.  The idea is that
pairs of particles do not interact unless they are touching, at
which point they are strongly repulsive.  A second important
category is softer colloidal particles, which are typically charge
stabilized.  This means that ions disassociate from their surface,
leaving their surface slightly charged with the counterions in the
solvent, similar to polyelectrolytes.  The like-charged
particles repel each other, again preventing particles from getting
close enough to each other to feel the van der Waals attraction.
For hard-sphere-like particles, the control parameter is the volume
fraction:  the fraction of volume occupied by the particles, which
of course is proportional to the particle concentration
\cite{poon12}.  For softer particles, the control parameter is
the concentration or number density \cite{royall13}.  To avoid
confusion, this Viewpoint will use the word concentration to refer
to the control parameter for colloidal samples.  Glasses are found
when the concentration is above the glass transition concentration,
where that specific concentration depends on the sample details.

\begin{figure}
\includegraphics[width=8cm]{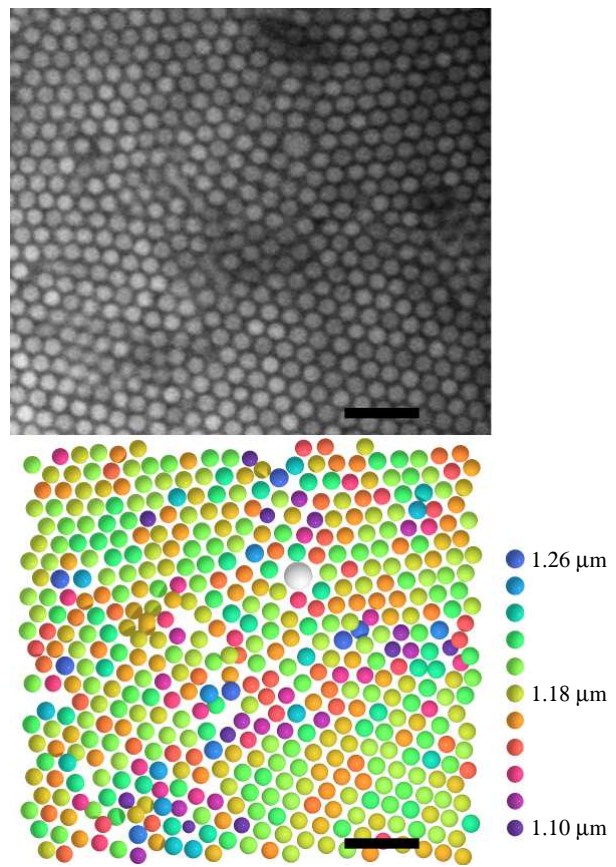}
\caption{Top:  Confocal microscope image of a colloidal crystal.
Bottom:  Rendered image of the same data, with the particles
colored by their size, with the anomalous large particle shaded
white.  The legend indicates how the color corresponds
to the particle radius.
The particles have
a mean radius of 1.18~$\mu$m and are drawn to scale.
While the polydispersity is only 0.045,
the particles that are smaller or larger than average tend to
cluster in more disordered regions.  In both panels, the scale
bar is 10~$\mu$m and drawn at the same location in the sample.
The sample was imaged in 3D and the rendered data are taken
from a region of thickness 2.2~$\mu$m; not all the particles are
perfectly co-planar.
The data are from Refs.~\cite{weeks00,kurita12} and have volume
fraction $\phi=0.46$.
}
\label{crystal}
\end{figure}

One other important consideration is the polydispersity of a
colloidal suspension.  Much like polymers, a batch of colloidal
particles will have a range of sizes.  The polydispersity is defined
as the standard deviation of particle sizes divided by the mean
size, using a number average; typical values are 5-8\%.  Samples
with a low polydispersity can organize into crystals
\cite{pusey86,henderson98,schope07,pusey09}, which is interesting in
its own right \cite{pusey89b,cheng01,leunissen05,jensen13}.  
Figure \ref{crystal} shows an image taken within a
colloidal crystal; the color indicates the relative
particle size.  The crystalline regions tend to have mostly similar-sized
particles, highlighting the importance of polydispersity.  Often
experimentalists who wish to study glass transition phenomena
will use more highly polydisperse samples, or else a bidisperse
mixture such as that shown in Fig.~\ref{narumipic}, much as is
done in simulations \cite{pusey09,miyagawa88,kob95a}.

\begin{figure}
\includegraphics[width=8cm]{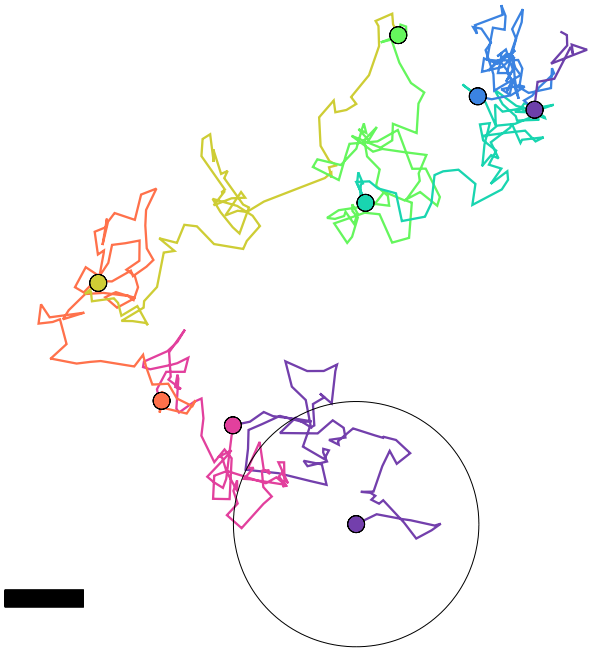}
\caption{$8\tau_D = 155$~s duration trajectory of a
colloidal particle with radius
$a=1.55$~$\mu$m and diffusivity
$D=0.062$~$\mu$m$^2$/s.  Segments of duration
$\tau_D$ are indicated by color and separated by small filled circles.
The large circular outline indicates the particle size.  The
scale bar is 1~$\mu$m.  This is from an experiment with a low
particle concentration, far from the glass transition
concentration.
}
\label{diffusivity}
\end{figure}

Another important experimental consideration is the particle
size, which determines the particle diffusivity and therefore the
relevant time scales of an experiment.
Colloidal particles undergo Brownian motion due to thermal energy.
In a liquid-like sample (below the glass
transition concentration) Brownian motion allows the particles to
rearrange, and macroscopically these rearrangements are what 
allows the sample to flow.
The typical time scale for particles to diffuse their own radius is
given by
\begin{equation}
\tau_D = \displaystyle \frac{a^2}{2 D} = \displaystyle \frac{3 \pi
\eta a^3}{k_B T},
\label{diffusion}
\end{equation}
where $a$ is the particle radius, $D$ is the diffusion constant
\cite{einstein1905a,sutherland1905}, $\eta$ is the solvent viscosity,
$k_B$ is Boltzmann's constant, and $T$ is the temperature
\cite{hunter12rpp}.  An example of a diffusing particle is shown
in Fig.~\ref{diffusivity} where the particle's position is marked with small
filled circles at intervals of $\tau_D$.
For polystyrene particles in water, this time
scale ranges from $0.8-800$~ms for particles of radius $a =
100$~nm to $a=1$~$\mu$m, which is the size range one typically sees
for colloidal glass experiments.  The $a^3$ dependence of $\tau_D$
allows this time scale to vary dramatically as the particle size is
changed.  Fundamentally, this sets a ``clock speed'' for a colloidal
experiment.  For example, one might state that near the glass
transition the relaxation time scale grows by $10^5$, meaning that
the relaxation time scale is $10^5 \tau_D$, and one might wish to
use particles of a size such that $10^5 \tau_D$ isn't an
unreasonable amount of time to wait for an experiment to finish
\cite{vanmegen98,brambilla09}.
The ability to undergo Brownian motion on experimentally
reasonable time scales helps define the upper limit to colloidal
particle diameters ($\sim 10$~$\mu$m).  

While diffusion is also relevant for molecules in a small
molecule glass former or a polymers, colloids
also have hydrodynamic interactions due to the solvent
\cite{stickel05,morris09}.  Does this mean they're a poor
model for glasses and you should stop reading this Viewpoint?
Absolutely not.  Of course,
the solvent viscosity sets a viscosity
scale for a colloidal suspension, much as $\tau_D$ sets a time
scale
for a colloidal experiment. 
When measuring the growth of
the viscosity near the colloidal glass transition, one examines
this growth relative to the solvent viscosity \cite{cheng02}.
Explaining the full rheological behavior requires understanding
the hydrodynamic interactions \cite{stickel05,morris09}.  On the
other hand, the functional form of this viscosity growth in
colloidal samples resembles the viscosity growth in more traditional
glass-forming systems \cite{cheng02}.  The glass transition is less
a question about the rheological details of the liquid-like samples
(which depends on hydrodynamics for colloids)
and more a question about the dramatic increase in
the zero-frequency viscosity.  (The zero-frequency viscosity is
the viscosity one would measure at very long times in an experiment
with the smallest possible imposed stress.)  In this sense of
capturing the correct zero-frequency behavior, hydrodynamic
interactions are not a limitation of colloids as a model system.
One might also wonder if diffusive dynamics (as described in the
previous paragraph) are a limitation, but several simulations have
demonstrated that the long-time glassy dynamics are independent
of the short time dynamics \cite{scala07,berthier07,tokuyama07}.

In fact, the dominant physics is the steric interaction of the
colloidal particles:  for a particle to move, other
particles must move out of its way.  Steric interactions are
thought to be important for understanding liquids and glasses,
and for example hard spheres are a simple well-studied model of
atoms in liquids \cite{cohen59,bernal64,widom67}.  The importance
of steric interactions over other particle interaction details
helps explain why similar behaviors are seen in computational
glass models using Lennard-Jones particles, hard spheres, and
soft spheres; and why these simulation results match colloidal
experimental results with hard-sphere-like particles as well
as softer particles \cite{hunter12rpp}.  For that matter,
in polymer glasses, the crowding of nearby
monomers is quite important to understand their glassiness,
and the fact that some of the monomers are linked together
may be less crucial \cite{rothbaglay16}.  This is reinforced by the
observation that the glass transition temperature in polymers
is independent of molecular weight (above some minimum molecular
weight) \cite{fox54,santangelo98}.


Turning now to the glass state itself:  a
glass is out of equilibrium.  In general this is because the
relaxation time scales in a glassy material exceed the
experimental time scales.  However, the
properties of the sample do evolve with time, a process termed {\it aging}.
In polymers, this manifests as physical aging, where it is
observed that samples slowly become denser as time passes.  One
related consequence is that the gas permeability of a polymer glass
decreases as the sample ages (which can be problematic for gas
separation applications) \cite{hodge95,pfromm95,huang04}.
The concept is that polymers rearrange to find better-packed
configurations, thus decreasing the overall volume and closing some
of the gaps where previously gas molecules could squeeze between
\cite{park97}.
These changes in the glass become exponentially slower as the
sample ages; the amount of change one sees between 10 minutes
and 100 minutes after the glass is formed would be similar to the
amount of change one sees between 10 hours and 100 hours
\cite{struik77}.

\begin{figure}
\includegraphics[width=8cm]{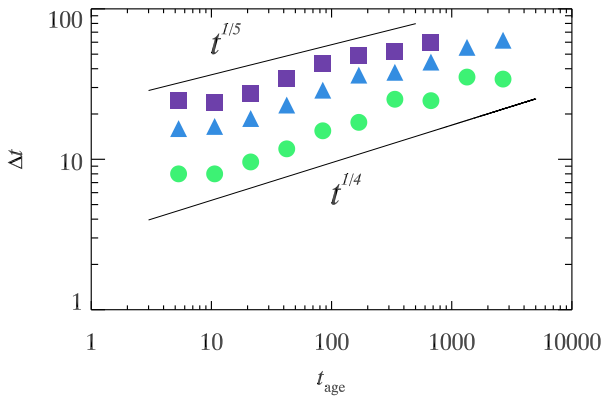}
\caption{
The time $\Delta t$ needed for particles to move a certain
distance as a function of the age of the sample.  Specifically
this is defined as $\langle | \vec{r}(t_{\rm age}+\Delta t) -
\vec{r}(t_{\rm age})|^2 \rangle = L^2$ where the angle
brackets are an average over all particles.  The values of
$L^2$ are 0.05, 0.10, and 0.20~$\mu$m$^2$ (circles,
triangles, and squares respectively).  For the square symbols at
large $t_{\rm age}$, the experiment concluded before 
the particles had diffused a distance $L^2$.
The sample is
composed of particles with a mean radius of 1.18~$\mu$m.  
The lines indicate power law growth with the exponents shown.
The data are from \cite{courtland03}.  
}
\label{agingfig}
\end{figure}

Likewise glassy colloids exhibit aging phenomena, in that
their properties slowly change with time.  Typically this is
examined by preparing a sample at a concentration such that it
is glassy, then shear-melting the sample by vigorous stirring.
After ending the stirring, the evolution of the sample is studied
\cite{courtland03,cipelletti03,peng14}.  This method is termed
``shear-rejuvenation.'' Alternatively, colloidal particles can be
used for which their size is temperature-controllable, and thus
temperature can be used to induce the particles to pack into a
glassy state \cite{di11,peng14,di14,peng16}.  This is more analogous to
the traditional temperature quench of a polymer glass.  With either
preparation protocol, aging of a colloidal glass is then seen as
particle motion slows with age.  Slight motions occur in the sample
(due to Brownian motion), and the time scale for these motions
grows as the sample ages as shown in Fig.~\ref{agingfig}.
This is quite similar to the slow evolution of aging polymer
samples \cite{struik77,hodge95}.  Experiments have shown that while
aging is seen following either preparation protocol, the details
of that aging differ between the two protocols even for the same
final conditions \cite{peng14}, as is also known to be the case
for polymer glasses \cite{mckenna03}.

However, in either case colloidal aging is observed at constant
concentration (constant volume), so this is distinctly different
from the physical aging of polymers.  What then does it mean for a
colloidal glass to age at constant volume?  One idea is that aging
is still the evolution of the sample toward a better packing of the
particles.  When the aging is initiated, the particles are in some
configuration set by the preparation protocol, but this is not the
equilibrium state.  Brownian motion still occurs, and occasionally
the particles rearrange in some way that brings them closer to an
ideal equilibrium state.  The closer the configuration is to the
ideal state, the lower the driving force is toward that equilibrium
state, and thus the dynamics should slow down.  Unfortunately, a
caveat is in order:  while this conceptual picture is sensible,
little data exist to support this story.  Attempts to observe structural
changes in aging colloidal samples have found few
\cite{kawasaki14} or no changes \cite{cianci06ssc,lynch08}.
Despite the scarce direct evidence,
this conceptual story must be true:  the sample has no
internal clock other than its structure, so the structure must
evolve as the sample ages.

\begin{figure}
\includegraphics[width=4.7cm]{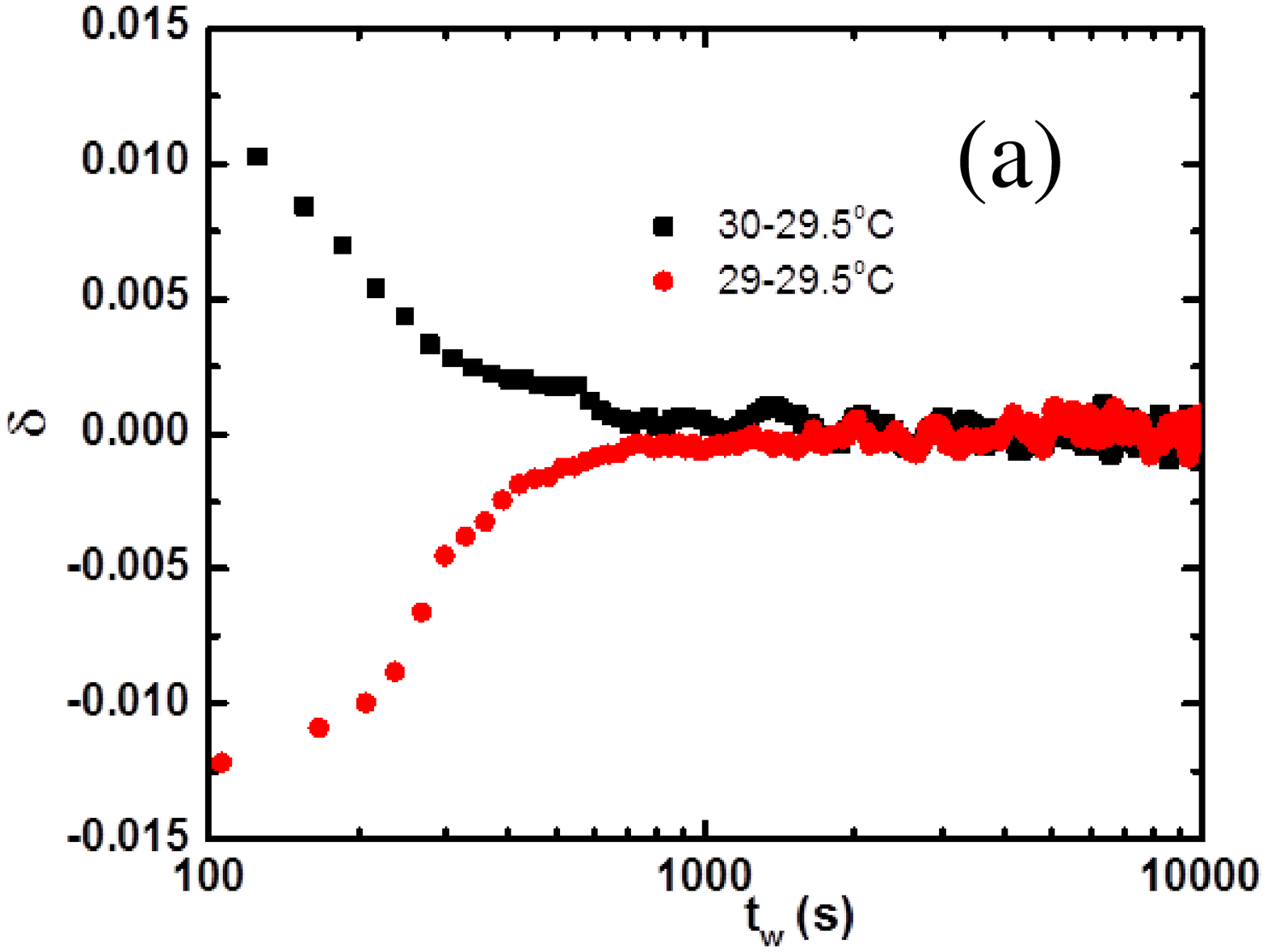}
\includegraphics[width=3.3cm]{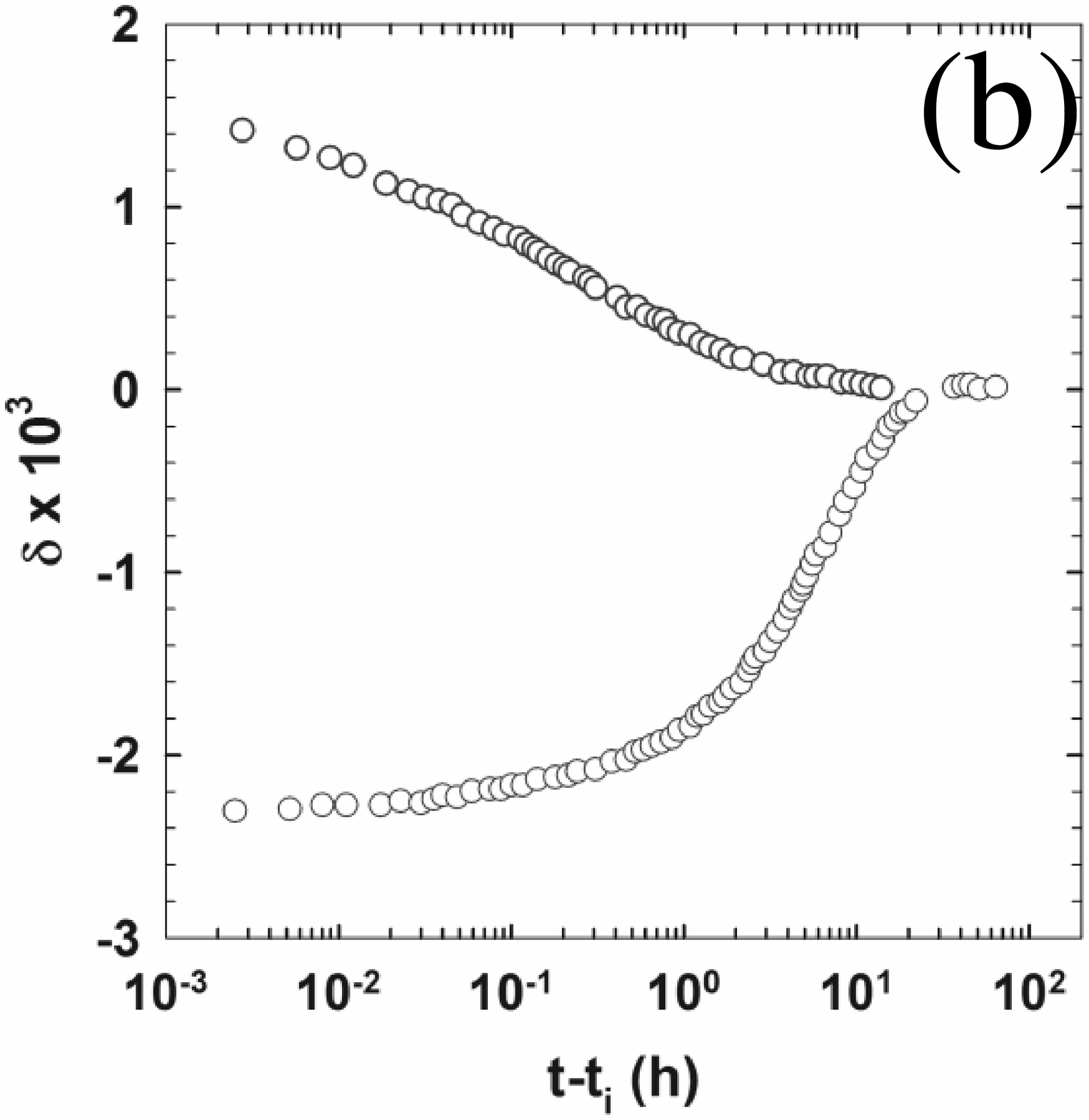}
\caption{
(a) Asymmetry of approach data from colloidal experiments using
temperature sensitive particles.  The temperature was adjusted
from the initial temperature (as shown) to a final temperature of
29.5$^\circ$~C.  $\delta$ is termed the departure from
equilibrium and is a measure of the
out-of-equilibrium dynamics.  (b) Asymmetry of approach data from
polymer glass experiments by Kovacs \cite{kovacs64}, as replotted
by Zheng and McKenna \cite{zheng03}.  A similar protocol was
followed with temperature, with both experiments set to the same
final temperature of 35$^\circ$~C.  Here $\delta$ is a measure of
the out-of-equilibrium sample volume.
(a) Reproduced from Di, Peng, and McKenna, {\it J. Chem. Phys.}
{\bf 140}, 054903 (2014), with the permission of AIP publishing.
(b) Reproduced
from Zheng and McKenna, {\it Macromol.} {\bf 36}, 2387 (2003) \cite{zheng03}.
}
\label{mckennafig}
\end{figure}

The McKenna group has used the temperature-sensitive colloids
mentioned above to do a series of clever experiments on colloidal
aging \cite{di11,di14,peng16} that mimic classic experiments by
Kovacs \cite{kovacs64}.  One such experiment studied the ``asymmetry
of approach'' to the equilibrated glass state.  In this experiment,
the sample is prepared in a glassy state and allowed to age for some
time.  The sample conditions are then jumped to a different glassy
state, and then studied as the sample evolves toward equilibrium
at the new glassy state.  This protocol is done twice, once with
the initial state less glassy than the final state, and once with the
initial state more glassy than the final state (more glassy in the sense
of being at a higher concentration for the colloids, or at a lower
temperature for the polymer glass).  For the polymer glass, Kovacs
found that the sample that starts at a glassier state takes longer
to evolve toward the final equilibrium, as shown in
Fig.~\ref{mckennafig}(b) \cite{kovacs64,zheng03}.  This shows
that the dynamics depend not only on the final temperature, but
also on the structure and history of the glass.  The difference
in subsequent behavior between the initially less glassy and
initially more glassy samples is why this is termed an ``asymmetry
of approach.''  Surprisingly, for colloids this is not really
seen, as shown in Fig.~\ref{mckennafig}(a) for one sample.
The approach toward the equilibrated final state takes the same
time, found in two different colloidal samples by the McKenna
group \cite{di11,di14}.  Either the dynamics do not depend on
the glassiness of the sample, or else the equilibration behavior
is dominated by the final conditions rather than the initial
conditions.

Above I introduced $\tau_D$ (Eqn.~\ref{diffusion}) as the
time scale for particles to diffuse their own radius in
a dilute sample.  In a concentrated sample, it takes much
longer for particles to diffuse their own radius; this is the
slowing of dynamics that characterizes the approach to the glass
transition.  Likewise, the viscosity of colloidal samples grows
dramatically as the glass transition is approached.  Looking at
Eqn.~\ref{diffusion}, one might suspect that replacing the
solvent viscosity $\eta$ with the macroscopic sample viscosity
$\eta(c)$ (at a particular concentration $c$) would produce
the new diffusive time scale, and that the
slowing of diffusion is a simple consequence of the growing
viscosity.  This, however, is not the case in glassy materials
\cite{poon12,zondervan07,swallen11,mazza07,ediger00}.  This is referred
to as the breakdown of the Stokes-Einstein relation between
diffusion and viscosity as the glass transition is approached
\cite{oppenheim96,jung04,berthier05epl}.  Microscopically, this
is likely
due to {\it dynamical heterogeneity}.  At any given moment,
different regions within the sample have different relaxation time
scales (spatial dynamical heterogeneity) and at different moments
a given region has different dynamics (temporal heterogeneity).
Equivalently, diffusive motion has different instantaneous
magnitudes in different regions.  Slowing diffusion as the glass
transition is approached is not just the sample slowing down;
rather, diffusion takes place in a fundamentally different fashion.

\begin{figure}
\includegraphics[width=8cm]{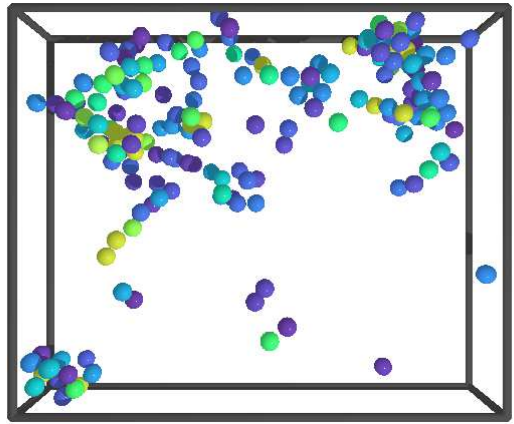}
\caption{
Rendering showing the positions of the most mobile colloidal
particles at a particular time for a sample close to the
colloidal glass transition.  The particles have
a radius of 1.18~$\mu$m and are drawn to scale.  The pictured
particles have displacements of at least 0.4~$\mu$m in the next
10~min, making them the top 5\% most mobile particles.  The color
indicates relative mobility, with the darker blue particles having
displacements $\sim 0.4$~$\mu$m and the lighter particles moving
as much as $\sim 1.0$~$\mu$m.  The box is $60 \times 50 \times
10$~$\mu$m$^3$.
The data are from Ref.~\cite{weeks00} ($\phi=0.52$ data).  
}
\label{dynhet}
\end{figure}

Simulations in the 1990’s first demonstrated dynamical
heterogeneity by visualizing which particles were
making large displacements at a given moment of time
\cite{hurley95,hurley96,kob97,poole98,donati98}.  A key
observation is that the particles with large displacements were
``cooperative'' in that neighboring particles moved in
similar directions \cite{donati98}.  For polymers, dynamical
heterogeneity is observable by adding in dye probes, or perhaps
grafting the dye probes to the polymer backbone
\cite{cicerone95,hall98,schob04}.  For example,
Ref.~\cite{schob04} used polarized light to observe the
fluorescence of single molecules and observed broad distributions
of rotational and translational correlation time scales.
For colloids, microscopy can be used to directly observe particle
motion in a sample.  In 1998 Kasper {\it et al.} first observed
dynamical heterogeneity of probe particles in dense colloidal
samples.  Using 2D samples, Marcus {\it et al.} could observe
all the particles in a region of the sample, and saw that
mobile regions were cooperative similar to what the simulations
had found.  Confocal microscopy allowed two different groups to
study dynamical heterogeneity in three-dimensional colloidal
samples \cite{kegel00,weeks00}, further confirming simulation
results.  Figure \ref{dynhet} shows an example taken from the data
of Ref.~\cite{weeks00}.  At the instant in time shown, the most
mobile particles are drawn, with the lighter colors indicating the
particles with the largest displacements.  The mobile particles
are clustered, leaving other regions with relatively immobile
particles at this instant.  At later times, different regions are
mobile and immobile.

\begin{figure}
\includegraphics[width=6cm]{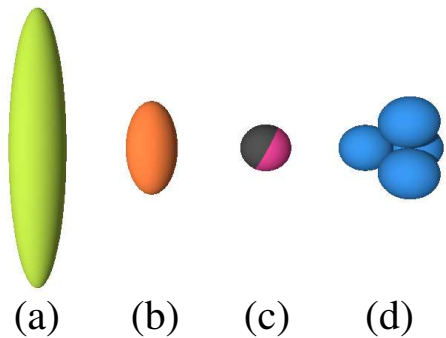}
\caption{
Illustrations of different particle shapes used in experiments to
observe rotational diffusion.
(a) Ellipsoid with aspect ratio 6 \cite{zheng11}.  (b) Ellipsoid with
aspect ratio 2 \cite{mishra13}.  (c) Sphere with optical
difference between two hemispheres \cite{kim11}.  (d) Tetrahedron
composed of four joined spheres \cite{edmond12pnas}.
}
\label{shapes}
\end{figure}

This discussion has focused on the {\it translational} diffusion of
particles from one location to another; 
recent advances in colloidal particle synthesis methods have enabled
striking observations of {\it rotational} diffusion.  These experiments
were motivated by prior experiments measuring rotational motion of probe
molecules in supercooled samples of polymers \cite{hall98} and 
small molecule liquids \cite{chang94}.  The Han
group synthesized colloidal ellipsoids with an aspect ratio of 6
[Fig.~\ref{shapes}(a)] 
and used sample chambers that confined these particles to a
quasi-two-dimensional layer, allowing for easy visualization of
translational and rotational motion of the ellipsoids
\cite{zheng11}.  At moderate concentrations, their particles
translate and rotate relatively easily.  At higher concentrations,
they found that the rotational motion underwent a glass transition,
but that particles could still translate.  At the highest
concentrations, both types of motion were glassy.  A related
experiment was published two years later, using a
quasi-two-dimensional layer and ellipsoids of aspect ratio two
[Fig.~\ref{shapes}(b)] 
\cite{mishra13}.  In this experiment, the two glass transitions
(rotation and translation) occurred at the same concentration.  In
both experiments, as the glass transition(s) were approached,
particles moved in cooperative groups -- that is, both rotational and
translational motion exhibited dynamical heterogeneity.  For the
large aspect ratio ellipsoids, the particles undergoing large
rotations were usually different from those undergoing large
translations \cite{zheng11}, whereas there were more particles
dually mobile for the smaller aspect ratio ellipsoids
\cite{mishra13}.

Two separate experiments examined rotational motion of tracers in
three-dimensional colloidal samples, and found opposite effects.
The first experiment used colloidal spheres which had been treated
so their orientation could be seen in a microscope image
[Fig.~\ref{shapes}(c)] 
\cite{kim11}.  These were added to samples of transparent spheres
close to the colloidal glass transition.  While their translational
motion slowed dramatically as the glass transition was approached,
rotational motion only slowed modestly.  In fact, their ratio
changed by a factor of 100 at the highest concentration they
studied, as compared to the dilute situation.  Were diffusion simply
a matter of the sample's macroscopic viscosity, this ratio
would have been independent of the concentration.  The second
experiment used tetrahedral clusters of spheres
[Fig.~\ref{shapes}(d)] also added to
samples of transparent spheres near the colloidal glass transition
\cite{edmond12pnas}.  In this situation, the observations were the
opposite of Ref.~\cite{kim11}:  both rotational and translational
diffusion slowed dramatically, and in this experiment it was the
rotational diffusion that was 50 times slower than the translational
diffusion at the highest concentration studied.  Both of these
experiments confirm that the decouping of translational and
rotational diffusion from each other occurs on the single
particle level, but they observe opposite directions of this
decoupling.

Despite observing opposite effects, both of these experiments can
be understood by recalling the basic physics discussed above.
The tetrahedral clusters of
Ref.~\cite{edmond12pnas} could sterically ``entangle'' with the
surrounding spheres.  Both their rotational and translational
degrees of freedom required the surrounding particles to move and
rearrange – which is the motion that is known to be dynamically
heterogeneous.  In contrast, the spherical tracers of
Ref.~\cite{kim11} interact sterically when they try to translate, but
interact hydrodynamically when they rotate.  That is, even if the
background particles were completely motionless, the spherical
tracers could still rotate, constrained only by a hydrodynamic drag
from the surrounding motionless particles
\cite{goldman67,imperio11}.  At higher concentrations, these
neighboring particles were closer to the tracers, so it is natural
that rotational motion slowed down; but this hydrodynamic effect was
far less significant than the steric hindrance of the translational
motion.  One can conjecture that upon gradually varying the probe
particle shape from a sphere to an ellipsoid, steric hindrance of
rotational motion would be gradually enhanced, and a crossover could
be seen from fast rotational dynamics to slow, glassy rotational
dynamics.  Theoretical and computational predictions suggest that
the aspect ratio of ellipsoids needs to be above some minimal
value \cite{demichele07}, for example 1.4 for 2D simulations
\cite{shen12}.  This prediction is also qualitatively consistent
with the ellipsoid experiments discussed above, where the aspect
ratio 6 ellipsoids had slower rotational dynamics \cite{zheng11}
and the aspect ratio 2 ellipsoids had identical translational and
rotational dynamics \cite{mishra13}.


Returning to broader questions about the scientific merits of
colloidal glasses, this
Viewpoint has argued that the colloidal glass transition is
a good model with many similarities to the glass transition of
polymers and small molecules.  Of course, one needs to be aware
of the advantages and disadvantages of any model.
A useful comparison is between colloidal experiments and
simulations.  In general, simulations of colloids are done to
understand situations where hydrodynamics are important
\cite{stickel05,morris09,brady88}.
Most typically these simulations aim to understand the
rheological behavior of colloidal suspensions at moderate
concentrations, and so the goal is not to understand the glass
transition.  Accordingly, there are relatively few simulations of
``the colloidal glass transition'' in comparison to the number of
simulations aimed at ``the glass transition.'' As noted
previously, there have been several studies that compared
realistic molecular dynamics at short time scales with Brownian
dynamics, all of which found that the short time scale dynamics
have no influence on the long time scale behavior of interest
\cite{scala07,berthier07,tokuyama07}.

The comparison to make, then, is the strengths and weaknesses of
colloidal glass transition experiments as compared to glass
transition simulations.  Experiments have the advantage of typically
having $10^8 - 10^{10}$ particles in a sample, allowing for
well-defined averages (when using light scattering) and at a minimum
avoiding finite size effects \cite{kim00,flenner15}.  Experiments
also study real materials which themselves might be of intrinsic
interest: toothpaste is essentially a colloidal glass, for example.
Simulations have the advantage that the particle interaction is
completely specified.  For example, even hard-sphere-like colloids
are not truly hard spheres and there are real challenges when
comparing them to hard sphere simulations \cite{royall13}.  Some
methods exist to measure pair-wise interactions in colloidal
experiments \cite{behrens01}, but one hopes that the conclusions
from an experiment are not too sensitive to the exact details of the
interparticle interactions.  A final advantage of simulations is
that certain useful tricks are easier with simulations, such as
reproducing initial conditions \cite{widmercooper04} or simulating
behavior in four spatial dimensions \cite{vanmeel09a,sengupta12}.

In the end, much progress has been made when simulations of
various types of particles agree with experimental results using
various types of colloidal particles which in turn agree with
experiments studying polymers or small molecules.  For example, dynamical
heterogeneity has been seen in Lennard-Jones simulations
\cite{kob97,donati98,berthier04}, hard particle simulations
\cite{doliwa00}, soft particle simulations \cite{hurley95,hurley96},
polymer simulations \cite{bennemann99}, hard-sphere-like colloids
\cite{kegel00,weeks00}, and soft colloids \cite{colin11} --
all of which complements experiments done with small molecule
glasses \cite{bohmer96,cicerone95otp,cicerone96} and polymer
glasses \cite{schmidt-rohr91,cicerone95,schob04}.  At this point
it is clear that the presence dynamical heterogeneity does not
depend on the system studied, and then each experiment or
simulation contributes to a larger picture.

There are indeed several large pictures of current interest.
As mentioned above, shape is an intriguing parameter to play with
for colloidal glasses, and there are many more shapes besides
simple clusters of spheres or ellipsoids \cite{glotzer07} which
may lead to a diversity of amorphous states \cite{damasceno12}.
Using complex shapes can lead to better understanding of how steric
interactions determine the glassiness of small molecule glasses.
Another current topic of interest is clarifying how packing problems
(especially of athermal particles) may relate, or not, to the
glass transition problem \cite{ikeda12}.  Since the late 90's
there was a conjecture that these problems were closely related
\cite{liu98}; recent simulations suggest that the similarities
are more superficial than had been thought \cite{ikeda12,ikeda13}.
Colloidal glass experiments by Basu {\it et al.} support the simulation
results \cite{basu14}, but questions remain how packing
structures and dynamics differ between thermal and athermal systems.
To mention a final topic, simulations and theories of the glass
transition often consider physically implausible situations
that lead to interesting insights, such as freezing a subset
of particles and observing how nearby particles are affected
\cite{biroli08,cammarota12,cammarota13prl}.  Recent experiments use
2D colloidal systems and holographic laser tweezers to duplicate
some of these conditions \cite{gokhale14,himanagamanasa15},
confirming many of the predictions.  Given continuing advances in
colloidal synthesis techniques \cite{glotzer07} and other clever
experimental techniques, it is likely that fruitful conversations
will long continue between those interested in colloidal glasses
and those interested in other types of glasses.



The author thanks C.~Cao, C.~B.~Roth, and S.~Vivek for helpful
discussions, and thanks G.~McKenna for providing
Fig.~\ref{mckennafig}.  This work was supported by the National Science
Foundation (DMR-1609763).

\bibliography{eric}

\end{document}